\magnification1200
\font\small=cmr7

%%%%%%%%%%%%%%%%%%%%%%%%%%%%%%%%%%%%%%%%%%%
%%%%%%%%%%%%%% the metrics %%%%%%%%%%%%%%%%
%%%%%%%%%%%%%%%%%%%%%%%%%%%%%%%%%%%%%%%%%%%

\vsize = 8.5truein
\hsize = 6truein %%%%% text width
\voffset = 5truemm %%%%% ver printing
\hoffset=-1mm
\baselineskip = 12 pt %%%%% line spacing
\nopagenumbers
\headline={\hfil}

%%%%%%%%%%%%%%%%%%%%%%%%%%%%%%%%%%%%%%%%%%%

%%%%%%%%%%%%%%%%%%%%%%%%%%%%%%%%%%%%%%%%%%%
%%%%%%%%%%%%%%% numberings %%%%%%%%%%%%%%%%
%%%%%%%%%%%%%%%%%%%%%%%%%%%%%%%%%%%%%%%%%%%

\newcount\ch %%%%% ch(apters
\newcount\eq %%%%% eq(uations
%\newcount\foo %%%%% foo(tnotes
\newcount\ref %%%%% ref(erences

\def\chapter#1{
\parag\eq = 1\advance\ch by 1{\bf\the\ch.\enskip#1}
}
\def\equation{
\eqno(\the\ch.\the\eq)\global\advance\eq by 1
}
\def\reference{
\parag [\number\ref]\ \advance\ref by 1
}
\ch = 0 %%%%% global init ch(apter
%%%%% eq is set to 1 by \chapter
%\foo = 1 %%%%% global init foo(tnote
\ref = 1 %%%%% global init ref(erence

%%%%%%%%%%%%%%%%%%%%%%%%%%%%%%%%%%%%%%%%%%%
%%%%%%%%%%%%%%%% the text %%%%%%%%%%%%%%%%%
%%%%%%%%%%%%%%%%%%%%%%%%%%%%%%%%%%%%%%%%%%%
\centerline{{\bf NON-RELATIVISTIC SUPERSYMMETRY}\footnote{*} {Walifest Ð MRST15 : New directions in the application of symmetry principles to elementary particle physics. Syracuse (N.Y.) 1993. Ed. J. Schechter. p.109-118. Singapoe : World Sientific (1994).}}
\vskip 1.truecm
\centerline{C. DUVAL}
\vskip 12truept
\centerline{ 
D\'epartement de Physique, Universit\'e d'AIX-MARSEILLE II (France).}
\vskip 0.5truecm
\centerline{and}
\vskip 0.5truecm
\centerline{P. A. HORV\'ATHY}
\vskip 12truept
\centerline{D\'epartement de Math\'ematiques, Universit\'e de TOURS (France).}
\vskip 0.7truecm
\noindent{\bf Abstract} {\it The supersymmetric extensions of
the Schr\"odinger algebra are reviewed.}
\vskip 0.3truecm

%%%%%%%%%%%%%%%%%%%%%%%%%%%%%%%%%%%%%%%%%%%%%%%%%%%%%%
%%%%%%%%%%%%% lower case greek boldface %%%%%%%%%%%%%%
%%%%%%%%%%%%%%%%%%%%%%%%%%%%%%%%%%%%%%%%%%%%%%%%%%%%%%

\font\tenb=cmmib10 %%%%%
\newfam\bsfam

\textfont\bsfam=\tenb

\mathchardef\betab="080C
\mathchardef\xib="0818
\mathchardef\omegab="0821
\mathchardef\deltab="080E
\mathchardef\epsilonb="080F
\mathchardef\pib="0819
\mathchardef\sigmab="081B
\mathchardef\bfalpha="080B
\mathchardef\bfbeta="080C
\mathchardef\bfgamma="080D
\mathchardef\bfomega="0821
\mathchardef\zetab="0810

%%%%%%%%%%%%%%%%%%%%%%%%%%%%%%%%%%%%%%%%%%%%%%%%%%%%%%
%%%%%%%%%%%%%%%%%% some definitions %%%%%%%%%%%%%%%%%%
%%%%%%%%%%%%%%%%%%%%%%%%%%%%%%%%%%%%%%%%%%%%%%%%%%%%%%

\def\parag{\hfil\break}
\def\and{\qquad\hbox{and}\qquad}

\def\kikezd{\parag\underbar}

\def\osp{\math{osp}}
\def\sch{\math{sch}}
\def\o{\math{o}}

\def\IR{{\bf R}}

\def\bQ{{\bf Q}}

\def\br{{\bf r}}
\def\bp{{\bf p}}
\def\bq{{\bf q}}

\def\bA{{\bf A}}
\def\bD{{\bf D}}
\def\bG{{\bf G}}
\def\bP{{\bf P}}
\def\cJ{{\cal J}}
\def\cB{{\cal G}}
\def\cG{{\cal G}}
\def\cP{{\cal P}}
\def\cM{{\cal M}}
\def\cD{{\cal D}}
\def\cK{{\cal K}}
\def\cL{{\cal L}}
\def\cH{{\cal H}}
\def\cQ{{\cal Q}}
\def\cS{{\cal S}}
\def\cX{\Xi}

\def\smallcirc{{\raise 0.5pt \hbox{$\scriptstyle\circ$}}}
\def\smallover#1/#2{\hbox{$\textstyle{#1\over#2}$}}
\def\2{{\smallover 1/2}}
\def\semidirectproduct{
{\ooalign
{\hfil\raise.07ex\hbox{s}\hfil\crcr\mathhexbox20D}}}
\def\ccr{\cr\noalign{\medskip}}
\def\={\!=\!}

\def\D{D\mkern-2mu\llap{{\raise+0.5pt\hbox{\big/}}}\mkern+2mu}

%%%%%%%%%%%%%%%%%%%%%%%%%%%%%%%%%%%%%%%%%%%%%%%%%%%%%%
%%%%%%%%%%%%%%% some other definitions %%%%%%%%%%%%%%%
%%%%%%%%%%%%%%%%%%%%%%%%%%%%%%%%%%%%%%%%%%%%%%%%%%%%%%

\def\math#1{\mathop{\rm #1}\nolimits}

%%%%%%%%%%%%%%%%%%%%%%%%%%%%%%%%%%%%%%%%%%%%%%%%%%%%%%
%%%%%%%%%%%%%%%%%%%%%%%%% main %%%%%%%%%%%%%%%%%%%%%%%
%%%%%%%%%%%%%%%%%%%%%%%%%%%%%%%%%%%%%%%%%%%%%%%%%%%%%%

\chapter{Non-relativistic Chern-Simons theory.}

The $c\to\infty$ limit of the Chern-Simons theory of a complex scalar field 
$\Phi$, interacting,
through the Chern-Simons term,
 with a scalar and a vector potential, $A_0$ and $\bA$, leads 
to the non-relativistic Lagrangian [1]
$$\eqalign{
\int d^2\br\left\{
{\kappa\over2c}(\partial_t\bA)\times\bA
-A_0\big[\kappa B+e|\Phi|^2\big]\right.
&+i\Phi^*\partial_t\Phi
\cr
&\left.-{1\over2m}|\bD\Phi|^2
+{e^2\over2mc\kappa}|\Phi|^4
\right\}.
\cr
}
\equation
$$
This theory is invariant with respect to the Galilei 
group, and has the following conserved quantities:
$$\left\{
\matrix{
\br\to\br+{\bf a}\hfill
&\bP=\displaystyle\int d^2\br
\underbrace{{1\over2i}\left[\Phi^*\bD\Phi-(\bD\Phi)^*\Phi\right]}_
{\bf\cal P}\;\hfill
&\hbox{momentum}\hfill
\ccr
\br\to A\br\hfill
&J=\displaystyle\int d^2\br\,\br\times{\bf\cal P}\hfill
&\hbox{angular momentum}\hfill
\ccr
\br\to\br+{\bf V}t\;\hfill
&\bG=t\bP-m\displaystyle\int d^2\br\,
\br\rho\hfill
&\hbox{galilean boost}\hfill
\ccr
t\to t+e\hfill
&H=\displaystyle\int d^2\br\left[\displaystyle{|\bD\Phi|^2\over2m}
-\lambda_1\rho^2\right]\hfill
&\hbox{energy}\hfill
\cr
}\right.
\equation
$$
where $\rho\=|\Phi|^2$ and $\lambda_1\=e^2/2mc\kappa$. It also has the 
non-relativistic 
\lq conformal' symmetry discovered in the early seventies by Niederer and
by Hagen [2], namely
\goodbreak
$$\left\{
\matrix{
\pmatrix{\br\cr t}\to
\pmatrix{d \cdot\br\cr d^2 \cdot t\cr}\hfill
&D=tH-\2\displaystyle\int d^2\br\,
\br\cdot{\bf\cal P}\hfill
&\hbox{dilatation}\hfill
\ccr
\pmatrix{\br\cr t\cr}\to\pmatrix{\displaystyle{\br\over1-ft}\ccr
\displaystyle{t\over1-ft}\hfill\cr}\;\hfill
&K=-t^2H+2tD+\2 m\displaystyle\displaystyle\int d^2\br\,
\br^2\rho\quad\hfill
&\hbox{expansion}\hfill
\cr
}\right.
\equation
$$

The $8$ generators in Eq. (1.2-3) form the planar {\it Schr\"odinger 
algebra} denoted by $\sch(2)$; time translations, dilatations and 
expansions span an $\o(2,1)$ subalgebra.

By taking the 
$c\to\infty$ limit of 
the $N\=2$ supersymmetric theory of Lee, Lee, and 
Weinberg [4], Leblanc, Lozano and Min [3] generalized (1.1) to
$$
\eqalign{
\int d^2\br\left\{{\kappa\over2c}\right.(\partial_t&\bA)
\times\bA-A_0\big[\kappa B+e(|\Phi|^2+|\Psi|^2)\big]
+i\Phi^*\partial_t\Phi+i\Psi^*\partial_t\Psi
\cr
&\left.-{1\over2m}\Big(|\bD\Phi|^2+|\bD\Psi|^2\Big)
+\displaystyle{e\over2mc}B|\Psi|^2
+\lambda_1|\Phi|^4+\lambda_2|\Phi|^2|\Psi|^2
\right\},
}
\equation
$$
where $\Psi$ is a two-component fermion field,  
$\lambda_1\=e^2/2mc\kappa$, and
$\lambda_2\=3\lambda_1$. (Notice here the Pauli term).
The extended system (1.4) is also Schr\"odinger-symmetric:
Putting $\rho_B\=|\Phi|^2$ and
$\rho_F\=|\Psi|^2$, one finds the conserved charges

$$\left\{
\matrix{
\bP=\displaystyle{\int} d^2\br
\underbrace{
{1\over2i}\left[\Phi^*\bD\Phi-(\bD\Phi)^*\Phi
+\Psi^*\bD\Psi-(\bD\Psi)^*\Psi\right]}_
{\bf\cal P}\;\hfill
&\hbox{momentum}\hfill
\ccr
J=\displaystyle\int d^2\br\,\big[\br\times{\bf\cal P}+\2\rho_F\big]\hfill
&\hbox{angular momentum}\hfill
\ccr
\bG=t\bP-m\displaystyle\int d^2\br\,
\br\big[\rho_B+\rho_F\big]\hfill
&\hbox{galilean boost}\hfill
\ccr
H=\displaystyle\int d^2\br
\Bigg[\displaystyle{1\over2m}\Big(|\bD\Phi|^2+|\bD\Psi|^2\Big)\hfill&
\cr
\qquad\qquad\qquad\qquad-\displaystyle{e\over2mc}B\rho_F
-\lambda_1\rho_B^2-\lambda_2\rho_B\rho_F\Bigg]\hfill
&\hbox{energy}\hfill
\ccr
D=tH-\2\displaystyle\int d^2\br\,
\br\cdot{\bf\cal P}\qquad\hfill
&\hbox{dilatation}\hfill
\ccr
K=-t^2H+2tD+\2 m\displaystyle\displaystyle\int d^2\br\,
\br^2\big(\rho_B+\rho_F\big)\qquad\hfill
&\hbox{expansion}\hfill
\cr
}\right.
\equation
$$
The theory is also invariant with respect to global rotations of 
 bosons and fermions, $\Phi\to e^{i\alpha}\Phi$ and 
$\Psi\to e^{i\beta}\Psi$, yielding the two central charges
$$\left\{
\matrix{
M_B=m\displaystyle\int d^2\br \rho_B\qquad\hfill
&\hbox{bosonic mass}
\ccr
M_F=m\displaystyle\int d^2\br \rho_F\hfill
&\hbox{fermionic mass}
\cr}\right.
\equation
$$

Leblanc et al. [3] demonstrate furthermore that this system has an 
$N\=2$ supersymmetry:
$$
Q_2=\displaystyle{1\over\sqrt{2m}}\,\displaystyle{\int}d^2\br\,\Phi^*D_+\Psi
\equation
$$
(where $D_+\=D_1+iD_2$) and its complex conjugate, $Q_2^*$, satisfy
$$
\{Q_2,Q_2^*\}=-iH.
\equation
$$
Adding 
$$
Q_1=i\sqrt{2m}\,\displaystyle{\int}d^2\br\,\Phi^*\Psi
\equation
$$
and its conjugate $Q_1^*$ 
(which satisfy $\{Q_1,Q_1^*\}=-2iM$), 
as well as
$$
F=i[K,Q_2]
\and
F^*=i[K,Q_2^*],
\equation
$$
one gets a closed algebra:
 $$
 \bP,J,\bG,H,D,K,M_B,M_F, Q_1,Q_1^*,Q_2,Q_2^*,F,F^*
$$ 
span a $16$-dimen\-sional supersymmeric 
ex\-ten\-sion of the Schr\"odinger algebra [3].

\chapter{Schr\"odinger superalgebras.}

Supersymmetric extensions of the Schr\"odinger algebra were considered 
previously [5].  
For example, Beckers et al. consider an $n$-dimensional fermionic harmonic 
oscillator with the
total Hamiltonian
$$
H_{\rm tot}=H_B+H_F={1\over2}\Big({\bp^2\over m}+m\omega^2\br^2\Big)
+\2\omega\sum_{a=1}^n\Big(\zeta_+^a\zeta_-^a-\zeta_-^a\zeta_+^a\Big),
\equation
$$
the $\zeta^a_\pm$
($a=1,\ldots,n)$ being the generators of a Clifford algebra. As 
pointed out by Niederer twenty years ago [6] for a bosonic oscillator and 
extended to the fermionic case by Beckers et al., $H_{\rm tot}$ admits 
the same Schr\"odinger symmetry as a free particle, with generators 
$$
\left\{
\eqalign{
J_{ab}&=r_ap_b-r_bp_a-({\zeta_+}_a{\zeta_-}_b+{\zeta_-}_a{\zeta_+}_b)
\ccr
H_B&={1\over2}\Big({{\bf p}^2\over m}+m\omega^2 {\bf r}^2\Big)
\ccr
C_\pm&=
\pm{i\over2m}e^{\mp2i\omega t}\big({\bf p}\pm im\omega {\bf r}\big)^2
\ccr
P_\pm&=
\pm ie^{\mp i\omega t}\big({\bf p}\pm im\omega {\bf r}\big)
\ccr
M&=m
\cr
}\right.
\equation
$$
Here $H_B$, $C_+$ and $C_-$ generate an
$\o(2,1)$ algebra; the angular momentum $J$ generates 
$\math{o}(n)$.
The $P_{\pm}$ and $M$ span the $n$-dimensional Heisenberg algebra
$\rm{h}(n)$.

This system also has an $N=2$ conformal supersymmetry, with supercharges
$$
\left\{
\eqalign{
&Q_\pm= \big({\bf p}\mp
im\omega {\bf r}\big)\cdot{\zetab_\pm\over\sqrt{m}},
\ccr
&S_\pm=e^{\mp2i\omega t}
\big({\bf p}\pm im\omega {\bf r}\big)\cdot{\zetab_\pm\over\sqrt{m}},
\ccr
&{\bf T}_\pm=
e^{\mp i\omega t}{\sqrt{m}}\ \zetab_\pm.
\cr
}\right.
\equation
$$
Note that $H_B$ and $H_F$ are both bosonic and are
separately conserved. 
In the plane ($n\=2$) the Beckers et al. algebra has $18$ generators; its 
 global structure is
$$
{\widetilde{\sch}}(2/2)\cong\Big(\o(2)\times\osp(1/2)\Big)
\,\semidirectproduct\,{\rm h}(n/2).
\equation
$$

The relation with the superalgebra of Leblanc et al. is explained in our recent paper 
[7]:

\kikezd{Theorem}. \hskip1mm 
{\it In any dimension $n$ and for any given integer $N$, the $n$-dimensional Schr\"odinger algebra 
$\sch(n)$  admits an $N$-supersymmetric extension with generators
\goodbreak
$$
\left\{
\matrix{
\cJ^{ab}\hfill
&=&Q^aP^b-P^aQ^b-\sum_{j=1}^N\xib_j^a\xib_j^b,\qquad
\qquad\hfill
&\cQ_j\hfill &=&\bP\cdot\xib_j,\hfill
\ccr
\cH\hfill &=&\2 \bP^2,\hfill
&\cS_j\hfill &=&\bQ\cdot\xib_j,\hfill
\ccr
\cD\hfill &=&\bP\cdot\bQ,\hfill
&{\bf T}_j\hfill &=&\xib_j,\hfill
\ccr
\cK\hfill &=&\2\bQ^2,\hfill&
\ccr
\cB\hfill &=&\bQ,\hfill&
\ccr
\cP\hfill &=&\bP,\hfill&
\ccr
\cM\hfill &=&1,\hfill
&\ccr
\cH_{jk}&=&\xib_j\cdot\xib_k,\hfill&
\cr
}\right.
\equation
$$
where $j,k\=1,\ldots,N$, $\bQ$, $\bP\in\IR^n$, and 
$\xib\in\IR_1^n$ is an $n$-dimensional Grassmann vector. The global 
structure of the algebra is that of
$$
\widetilde{\sch}(n/N)\cong\Big(\o(n)\times\osp(1/N)\Big)
\,\semidirectproduct\,{\rm h}(n/N).
\equation
$$

\goodbreak
For $n\!\neq\!2$ the extension is  unique.
For $n\=2$, however, the system admits an additional \lq twisted' extension: 
For any integer $\nu$, this latter is given by
the generators}
\goodbreak
$$\left\{
\matrix{
\cJ&=&
\bQ\times\bP,\qquad\hfill
&\cQ_j&=&
\bP\cdot\xib_j,\hfill
\ccr
\cH&=&\2\bP^2,\hfill
&\cQ^*_j&=&
\bP\times\xib_j,\hfill
\ccr
\cD&=&
\bP\cdot\bQ,\hfill
&\cS_j&=&
\bQ\cdot\xib_j,\hfill
\ccr
\cK&=&\2\bQ^2,\hfill
&\cS^*_j&=&
-\bQ\times\xib_j,\hfill
\ccr
\cG&=&\bQ,\hfill
&{\bf T}_j&=&\xib_j,\hfill
\ccr
\cP&=&\bP,\hfill
&
\ccr
\cM&=&1,\hfill
&
\ccr
\cH_{jk}&=&\xib_j\cdot\xib_k,\hfill
&
\ccr
\cL_{jk}&=&\xib_j\times\xib_k.\qquad
\hfill
&
\cr
}\right.
\equation
$$
where $j\=1,\ldots,\nu$. Note that,
in the plane, the cross product of two vectors is a
scalar, ${\bf u}\times{\bf v}=\varepsilon_{ij}u^iv^j$
where $\varepsilon_{ij}$ is the
totally antisymmetric symbol,
$\varepsilon_{12}=1$. 

The fermionic charges 
(with the exception of supertranslations) come in pairs, 
one for the scalar
product and one for the cross product, respectively. Also, there are two
rather then just one types of \lq fermionic Hamiltonians', 
namely $\cH_{jk}$ and 
$\cL_{jk}$.

As explained in Ref. 7, such  \lq twisted' extensions can only arise in the plane. The clue is that that the internal rotations in $\osp$ have to 
commute with ordinary space rotations, and this only happens in two dimensions. Put in another way, it is only in the plane that we can have two 
\lq scalar' products, one symmetric, the other antisymmetric.

For $\nu\=1$ there is no $\cH_{jk}$-type charge, and our 
\lq twisted' super-schr\"odinger algebra has the structure of
$$
\widetilde{\sch}_e(1)\cong\Big(\math{o}(2)\times\osp(1/2)\Big)
\,\semidirectproduct\,\rm{h}(2/1),
\equation
$$
which is again an $N\=2$ supersymmetric extension of the planar 
Schr\"odinger 
algebra.
The difference with (2.4) is that there is now just one, rather than 
two Grassmann variables and hence one, rather than two
super-translations.

Physical examples of this supersymmetry are provided by 
the magnetic vortex [11] presented in the next Section, 
and by Chern-Simons theory [3].
The dictionary between the super\-algebra of Leblanc
et al. [3] and our twisted super-Schr\"odinger algebra $\widetilde{\sch}_e(1)$ is presented in
 Table I. here below.
 
\vskip3mm
\noindent
\underbar{\bf TABLE} I.\hskip2mm {\small The dictionary between our superalgebra (2.7) with that of Leblanc et al. [3].} 
$$
\matrix{
&\hbox{\underbar{LLM}}\qquad\qquad\hfill&&\qquad\hbox{\underbar{DH}}
\hfill\ccr
&J\hfill&=&\qquad\cJ+\cM-{\smallover1/4}\cL\hfill
\ccr
&H\hfill&=&\qquad\cH\hfill
\ccr
&K\hfill&=&\qquad\cK\hfill
\ccr
&D\hfill&=&\qquad\2\cD\hfill
\ccr
&G_\pm\hfill&=&\qquad\mp\cB_1+i\cB_2\hfill
\ccr
&P_\pm\hfill&=&\qquad\mp\cP_1+i\cP_2\hfill
\ccr
&N_F\hfill&=&\qquad-\2\cL\hfill
\ccr
&N_B\hfill&=&\qquad\cM+\2\cL\hfill
\ccr
&Q_1\hfill&=&\qquad\cX^1+i\cX^2\hfill
\ccr
&Q_1^*\hfill&=&\qquad-i(\cX^1-i\cX^2)\hfill
\ccr
&Q_2\hfill&=&\qquad\2(\cQ+i\cQ^*)\hfill
\ccr
&Q_2^*\hfill&=&\qquad-{\smallover i/2}(\cQ-i\cQ^*)\hfill
\ccr
&F\hfill&=&\qquad-{\smallover i/2}(\cS+i\cS^*)\;\hfill
\ccr
&F^*\hfill&=&\qquad\2(\cS-i\cS^*).\hfill
\ccr
}
$$

\chapter{Supersymmetry of the magnetic vortex.}

A couple of years ago Jackiw [8] pointed out that a spin-$0$
particle in a Dirac monopole field has an $\o(2,1)$ dynamical
symmetry, generated by the spin-$0$ Hamiltonian, $H_0\=\pib^2/2m$
where $\pib\=\bp-e{\bf A}$, by the dilatation and by the expansion,
$$
D\=tH_0-\smallover1/4\{\pib,\br\}
\quad
\hbox{and}\quad
K\=-t^2H_0+2tD+m\br^2/2,
\equation
$$ 
respectively, to which angular momentum adds an $\o(3)$.
This allowed him to calculate the spectrum and the wave functions
group-theoretically.

Jackiw's result was extended to spin-$\2$ particles by D'Hoker and
Vinet [9] who have shown that for the Pauli Hamiltonian
$$
H={1\over2m}\left[\pib^2-e{\bf B}\cdot\sigmab\right]
\equation
$$
not only the conformal generators $D$ and $K$, but also
the fermionic generators
$$
Q\={1\over\sqrt{2m}}\,\pib\cdot\sigmab
\quad\hbox{and}\quad
S\=\sqrt{m/2}\,\br\cdot\sigmab-tQ
\equation
$$
are conserved.
Thus the spin system admits an
$\o(3)\times\osp(1/1)$ conformal supersymmetry, providing us with an algebraic solution of the Pauli equation.

More recently, Jackiw [10] found that the $\o(2,1)$ symmetry,
generated by --- formally --- the same $D$ and $K$ as
above, is also present for a magnetic vortex (an idealization for
the Aharonov-Bohm experiment), allowing for a group-theoretic
treatment of the problem.

Very recently [11] we  pointed out that the
$N\=2$ supersymmetry of the Pauli Hamiltonian of a
spin-$\2$ particle (present for any magnetic field in the plane [12]) combines,
 for a magnetic vortex, with
Jackiw's $\o(2)\times\o(2,1)$ into an $\o(2)\times\osp(1/2)$
superalgebra, which is a sub-superalgebra of the twisted super-Schr\"odinger algebra (2.7). 
To see this, let us consider a spin-$\2$ particle in a static magnetic field
${\bf B}\=\big(0,0,B(x,y)\big)$.
Dropping the irrelevant $z$ variable, we can work in the plane.
Then our model is described by the Pauli Hamiltonian
which is the planar version of (3.2),
$$
H={1\over2m}\left[\pib^2-eB\sigma_3\right]\,,
\equation
$$

It is now easy to see that the Hamiltonian  is a perfect square in {\it two different} ways:
both operators
$$
Q={1\over\sqrt{2m}}\,\pib\cdot\sigmab
\and
Q^*={1\over\sqrt{2m}}\,\pib\times\sigmab
\equation
$$
where $\sigmab\=(\sigma_1,\sigma_2)$, satisfy
$$
\{Q,Q\}=\{Q^\star,Q^\star\}=2H.
\equation
$$
Thus for  any static and purely magnetic field in the plane, $H$ is an
$N\=2$ supersymmetric Hamiltonian. The `twisted' charge $Q^\star$
was used, e.g., by Jackiw [13], to describe the Landau levels
in a constant magnetic field --- a classic example of 
supersymmetric quantum mechanics [12] --- 
where the supercharge $Q$ is a standard object to be looked at. 

Let us assume henceforth that $B$ is the field of a point-like
magnetic vortex directed along the $z$-axis, $B\=\Phi\,\delta(\br)$,
where $\Phi$ is the total magnetic flux.
This can be viewed as an idealization of the spinning version 
of the Aharonov-Bohm
experiment [14].

Inserting 
$A_i(\br)\=-(\Phi/2\pi)\,\epsilon_{ij}\,\br^j/r^2$ into the Pauli
Hamiltonian $H$, it is  straightforward to check
that 
%$
%D=tH-\smallover1/4 \left\{\pib,\br\right\}
%$
%and
%$K=-t^2H+2tD+\2mr^2
%$
$D$ and $K$ in the planar version of (3.1)
generate, along with $H$,
the $\o(2,1)$ Lie algebra:
$$[D,H]\=-iH,\qquad
[D,K]\=iK,\qquad
[H,K]\=2iD.
\equation
$$
The angular momentum,
$J\=\br\times\pib$, adds to this $\o(2,1)$ an extra $\o(2)$.
(The correct definition of angular momentum requires boundary conditions [15]).

Commuting $Q$ and $Q^\star$ with the expansion, $K$,
yields two more generators, namely
$$\eqalign{
S&=i[Q,K]
=\sqrt{m\over2}\left(\br-{\pib\over m}t\right)\cdot\sigmab,
\ccr
S^\star
&=i[Q^\star,K]
=\sqrt{m\over2}\left(\br-{\pib\over m}t\right)\times\sigmab.
\cr}
\equation
$$

It is now straightforward to see that both sets $Q,S$ and
$Q^\star,S^\star$ extend
the $\o(2,1)\cong\osp(1/0)$ into an $\osp(1/1)$ superalgebra.
However, these two algebras do not close yet: the `mixed'
anticommutators $\{Q,S^\star\}$ and $\{Q^\star,S\}$ bring in a new
conserved charge, {\it viz.}
$$
\{Q,S^\star\}\=-\{Q^\star,S\}=J+2\Sigma,
\equation
$$
where $\Sigma=\2\sigma_3$.
$J$ satisfies non-trivial
commutation relations with the supercharges,
$$
[J,Q]=-iQ^\star,\qquad
[J,Q^\star]=iQ,\qquad
[J,S]=-iS^\star,\qquad
[J,S^\star]=iS.
\equation
$$
Thus, setting
$$
Y\=J+2\Sigma=\br\times\pib+\sigma_3,
$$
the generators $H,D,K,Y$ and
$Q,Q^\star,S,S^\star$ satisfy
$$
\matrix{
[Q,D]\hfill&=&\smallover i/2 Q,\hfill
&[Q^\star,D]\hfill&=&\smallover i/2 Q^\star,\hfill
\ccr
[Q,K]\hfill&=&-iS,\hfill
&[Q^\star,K]\hfill&=&-iS^\star,\hfill
\ccr
[Q,H]\hfill&=&0,\hfill
&[Q^\star,H]\hfill&=&0,\hfill
\ccr
[Q,Y]\hfill&=&-iQ^\star,\hfill
&[Q^\star,Y]\hfill&=&iQ,\hfill
\ccr
[S,D]\hfill&=&-\smallover i/2 S,\hfill
&[S^\star,D]\hfill&=&-\smallover i/2 S^\star,\hfill
\ccr
[S,K]\hfill&=&0,\hfill
&[S^\star,K]\hfill&=&0,\hfill
\ccr
[S,H]\hfill&=&iQ,\qquad\hfill
&[S^\star,H]\hfill&=&iQ^\star,\hfill
\ccr
[S,Y]\hfill&=&-iS^\star,\hfill
&[S^\star,Y]\hfill&=&iS,\hfill
\ccr
\{Q,Q\}\hfill&=&2H,\qquad\qquad\hfill
&\{Q^\star,Q^\star\}\hfill&=&2H,\hfill
\ccr
\{S,S\}\hfill&=&2K,\hfill
&\{S^\star,S^\star\}\hfill&=&2K,\hfill
\ccr
\{Q,Q^\star\}\hfill&=&0,\hfill
&\{S,S^\star\}\hfill&=&0,\hfill
\ccr
\{Q,S\}\hfill&=&-2D,\hfill
&\{Q^\star,S^\star\}\hfill&=&-2D,\hfill
\ccr
\{Q,S^\star\}\hfill&=&Y,\hfill
&\{Q^\star,S\}\hfill&=&-Y.\hfill
\cr}
\equation
$$

Added to the $\o(2,1)$ relations, this means that our generators
span the $\osp(1/2)$ superalgebra.
On the other hand, 
$$
Z\=J+\Sigma\=\br\times\pib+\2\sigma_3
\equation
$$
commutes with all generators of $\osp(1/2)$, so that the full
symmetry is the direct product
$\osp(1/2)\times\math{o}(2)$, generated by
\goodbreak
$$
\left\{
\matrix{
Y\hfill&=&
\br\times\pib+\sigma_3,\qquad\qquad\hfill
&Q\hfill
&=&\displaystyle{1\over\sqrt{2m}}\,\pib\cdot\sigmab,\hfill
\ccr
H\hfill&=&
\displaystyle{1\over2m}\,
\left[\pib^2-eB\sigma_3\right],\qquad\quad\hfill
&Q^\star\hfill &=&
\displaystyle{1\over\sqrt{2m}}\,\pib\times\sigmab,\hfill
\ccr
D\hfill &=&
-\smallover 1/4\,
\left\{\pib,\bq\right\}
-t\displaystyle{eB\over2m}\,\sigma_3,\quad\hfill
&S\hfill
&=&\sqrt{\displaystyle{m\over2}}\,\bq\cdot\sigmab,\hfill
\ccr
K\hfill&=&
\2m\bq^2,\hfill
&S^\star\hfill&=&
\sqrt{\displaystyle{m\over2}}\,\bq\times\sigmab,\hfill
\ccr
Z\hfill&=&
\br\times\pib+\2\sigma_3,\hfill&
\ccr
}\right.
\equation
$$
where we have put
$\bq\=\br-(\pib/m)t$. Notice that (3.6) is a sub-superalgebra 
of the quantized version of the \lq twisted' 
superalgebra (2.7) obtained by the substitutions 
$P\to\pib$, $Q\to\bq$ and $\xi\to\sigmab$.

\kikezd{Acknowledgements}. 
We would like to thank Martin Leblanc and Jean-Guy Demers
for discussions and correspondence.

%%-----
\vskip2mm

\centerline{\bf References}

\reference
C.~R. Hagen, Ann. Phys. {\bf 157}, 342 (1984); Phys. Rev. {\bf D1}, 848, (85); 
2135 (1985);
R.~Jackiw and S.-Y.~Pi, Phys. Rev. {\bf D42}, 3500 (1990);
Phys. Rev. Lett. {\bf 67}, 415 (1991);
Phys. Rev. {\bf D44}, 2524 (1991);
Z.~F.~Ezawa and M.~Hotta, Phys. Rev. Lett. {\bf 67}, 411 (1991).

\reference
U. Niederer, Helv. Phys. Acta {\bf45}, 802 (1972);
C. R. Hagen, Phys. Rev. {\bf D5}, 377 (1972).

\reference
M.~Leblanc, G.~Lozano and H.~Min,
Ann. Phys. (N.Y.) {\bf 219}, 328 (1992).

\reference
C.~Lee, K.~Lee and E.~Weinberg, Phys. Lett. {\bf 243}, 105 (1990).

\reference
J.~Beckers and V.~Hussin,
Phys. Lett. {\bf A118}, 319 (1986);
J.~Beckers, D.~Dehin and V.~Hussin,
J. Phys. {\bf A20}, 1137 (1987);
J.~P.~Gauntlett, J.~Gomis and P.~K.~Townsend,
Phys. Lett. {\bf B248}, 288 (1990).

\reference
U. Niederer, Helv. Phys. Acta {\bf 46}, 192 (1973).

\reference
C.~Duval and P.A.~Horv\'athy,
Marseille Preprint CPT-93/P.2912, published as
J.Math.Phys. 35 (1994) 2516-2538.

\reference
R.~Jackiw,
Ann.~Phys. (N.Y.) {\bf 129}, 183 (1980).

\reference E.~D'Hoker and L.~Vinet,
Phys.~Lett. {\bf 137B}, 72 (1984);
Comm.~Math.~Phys. {\bf 97}, 391 (1985).

\reference
R.~Jackiw,
Ann.~Phys. (N.Y.) {\bf 201}, 83 (1990);
A. O. Barut and R. Wilson,
Ann. Phys. {\bf 164}, 223 (1985).

\reference
C. Duval and P. A. Horv\'athy,
After our paper was submitted to Phys. Rev. Lett, we became aware of the papers 
C.J.~Park, Nucl. Phys. {\bf B376}, 99 (1992) and
J-G. Demers, Mod. Phys. Lett. {\bf 8}, 827 (1993), which contain similar results. It has therefore been consequently withdrawn and included into ref. [7].

\reference
E.~Witten,
Nucl.~Phys. {\bf B185}, 513 (1981);
P.~Salomonson and J.W.~Van Holten,
Nucl.~Phys. {\bf B169}, 509 (1982);
M.~De Crombrugghe and V.~Rittenberg, Ann.~Phys. (N.Y.)
{\bf 151}, 99 (1983).

\reference
R.~Jackiw,
Phys.~Rev. {\bf D29}, 2375 (1984).

\reference
C. R. Hagen, Phys. Rev. Lett. {\bf 64}, 503 (1990);
R.~Musto, L.~O'Raifeartaigh and A.~Wipf,
Phys. Lett. {\bf B175}, 433 (1986);
P. Forg\'acs, L.~O'Raifeartaigh and A.~Wipf,
Nucl. Phys. {\bf B293}, 559 (1987).

\reference
P.A.~Horv\'athy, Phys.~Rev. {\bf A31}, 1151 (1985);
F.~Wilczek, Phys.~Rev.~Lett. {\bf 48}, 1144 (1982);
R.~Jackiw and A.N.~Redlich, Phys.~Rev.~Lett. {\bf 50}, 555 (1983);
W.C.~Henneberger, Phys. Rev. Lett. {\bf 52}, 573 (1984).

\bye